\documentclass[epj,twocolumn]{webofc}
\usepackage[varg]{txfonts}   
\woctitle{Hadron Collider Physics symposium 2012}
\begin{document}
\hyphenation{meth-ods}
\hyphenation{using}
\hyphenation{i-ne-las-tic}
\hyphenation{se-ve-ral}
\hyphenation{con-ti-nu-ous}
\hyphenation{me-a-su-re-ment}
\hyphenation{co-o-pe-ra-ti-on}
\hyphenation{me-a-su-re}
\hyphenation{de-scri-bing}
\hyphenation{re-sul-ting}
\hyphenation{cha-ra-cte-ri-zed}
\hyphenation{re-so-lu-ti-on}
\hyphenation{in-te-ra-cti-on}
\hyphenation{a-na-ly-ses}
\hyphenation{a-na-ly-sis}
\hyphenation{mo-ving}
\hyphenation{bet-we-en}
\hyphenation{i-te-ra-ti-ve}
\hyphenation{me-a-su-red}
\hyphenation{fun-cti-on}
\hyphenation{the-o-rem}
\hyphenation{Fi-gu-re}
\hyphenation{ex-pe-ri-men-tal}
\hyphenation{pos-si-bi-li-ty}
\hyphenation{ge-ne-ra-tors}
\hyphenation{re-pre-sen-ting}
\title{Summary of Physics Results from the TOTEM Experiment}

\author{Giuseppe Latino\inst{1}\fnsep\thanks{\email{giuseppe.latino@pi.infn.it}}, 
  on behalf of TOTEM Collaboration.  
}

\institute{
   Department of Physics, Via Roma 56 - 53100 Siena.
   Universit\`{a} degli Studi di Siena and Gruppo Collegato INFN di Siena, Italy.
}
\abstract{
The TOTEM experiment has performed several measurements related to its physics program
in dedicated (high ${\beta}^*$, low $\mathcal{L}$) LHC fills at $\sqrt{s}$ = 7 TeV. Under various 
beam and background conditions, the differential elastic (as a function of |t|), elastic, 
inelastic and total pp cross-sections have been measured. A measurement of the forward 
charged particle $\eta$ density has also been performed.
A summary of these measurements is here reported, as well as the first results from 
runs at the LHC energy of $\sqrt{s}$ = 8 TeV. 
}
\maketitle
\section{Introduction}
\label{intro}
TOTEM \cite{TOTEM_TDR} is one of the six experiments that investigate the 
frontier of high energy physics at the LHC. It has been designed for a precise 
measurement of the total $pp$ cross-section ($\sigma_{tot}$) down to 1$\div$2\,$\%$, 
for the study of the nuclear elastic $pp$ differential cross-section ($d{\sigma}_{el} / dt$) over 
a wide range of the squared four-momentum transfer $|t| \sim (p\theta)^2$, as well as
for the study of inelastic and diffractive processes (to be performed partially in cooperation 
with CMS). In particular, TOTEM is able to measure $\xi$- ($\equiv \Delta p/p$), $|t|$- and mass- 
distributions for the intact (``leading'') protons produced in diffractive events, with
acceptances depending on the beam optics.

Given the lack of a complete description, based on a solid theoretical ground in the framework of 
the QCD, for the ``soft'' hadron interactions, many details of diffractive (due to color singlet 
exchange) and non-diffractive (due to color exchange) processes are still poorly understood. 
At the same time these processes, with close links to proton structure and low energy QCD, represent a 
big fraction of $\sigma_{tot}$. Consequently, the experimental measurements performed by TOTEM are 
very important because they represent a reference for the phenomenological models 
describing the soft hadron interactions. 

Furthermore, the study of the charged particle production in the forward region provides 
a significant contribution to the understanding of cosmic ray (CR) physics, heavily based on the 
simulation of primary CR particle interactions with the atmosphere. The existing 
models are indeed characterized by significant inconsistencies in the forward region for the 
predictions on energy flow, multiplicity and other quantities related to cosmic ray air showers.
Thanks to its forward trackers TOTEM is able to perform the measurements of the charged particle 
pseudorapidity\footnote{
The pseudorapidity $\eta$ is defined as: $\eta = -ln(tan\frac{\theta}{2})$, where $\theta$ is the 
polar angle of the particle direction with respect to the counterclockwise beam direction.
}
density (dN$_{ch}/d\eta$) and multiplicity distribution (N$_{ch}$) in a previously 
unexplored forward $\eta$ region.

Finally, a common data taking of TOTEM with CMS (resulting in the largest acceptance detector ever 
built at a hadron collider) opens the possibility of more detailed studies on inelastic processes, 
including hard diffraction \cite{CMS_TOTEM_TDR}.

After a brief description of the detector apparatus, a summary of the current TOTEM physics results 
is given in this paper.
\section{Detector setup}
\label{detector}
\begin{figure}[!htb]
\centering
\includegraphics[width=1.\linewidth,clip]{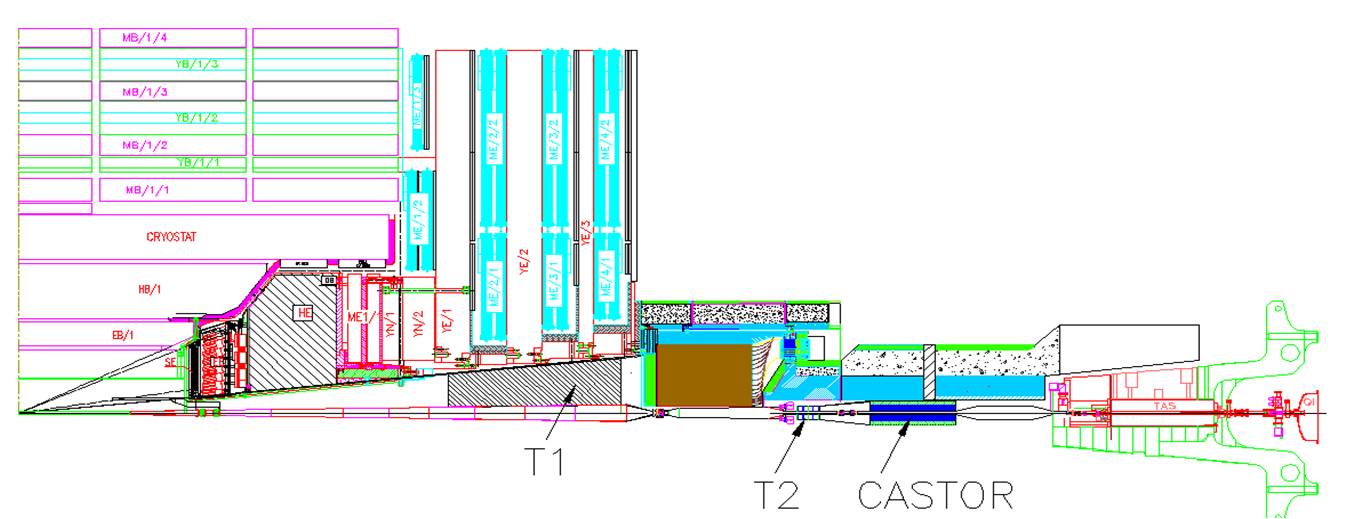}
\includegraphics[width=1.\linewidth,clip]{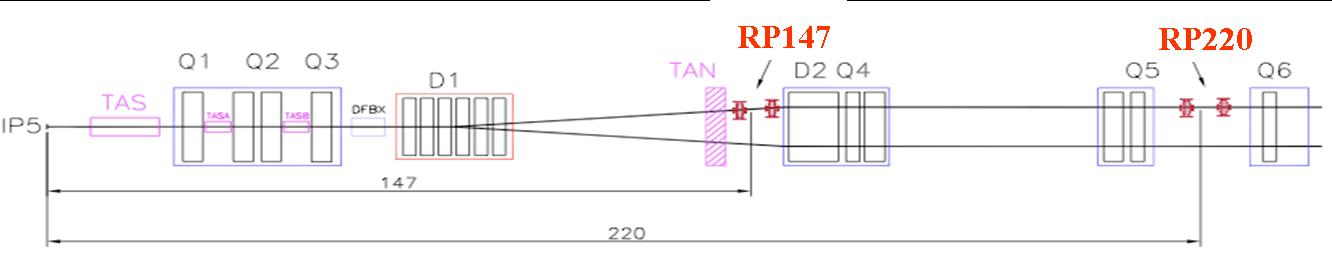}
\caption{Top: T1 and T2 location in the forward region of CMS.
Bottom: Roman Pots location along the LHC beam-line.
All TOTEM detectors are installed on both sides of IP 5.}
\label{fig:totem_exp}
\end{figure}

In order to fulfill its physics program TOTEM has to cope with the challenge of triggering
and recording events in the very forward region with a good acceptance for particles produced
at very small angles with respect to the beam. This involves the detection of
elastically scattered and diffractive protons at a location very close to the beam, together
with efficient forward charged particle detection in inelastic events with losses reduced to few
per-cents. This is accomplished with its experimental setup, characterized by three detector
components located on both sides of the interaction point (IP) 5, shared with CMS
(figure~\ref{fig:totem_exp}): the T1
and T2 inelastic telescopes, embedded inside the forward region of CMS, and the ``Roman Pots''
detectors, placed on the beam-pipe of the outgoing beams in two stations at about 147 m and
220 m from IP 5.

\begin{figure}[htb]
\centering
\includegraphics[width=0.49\linewidth,height = 3.5 cm]{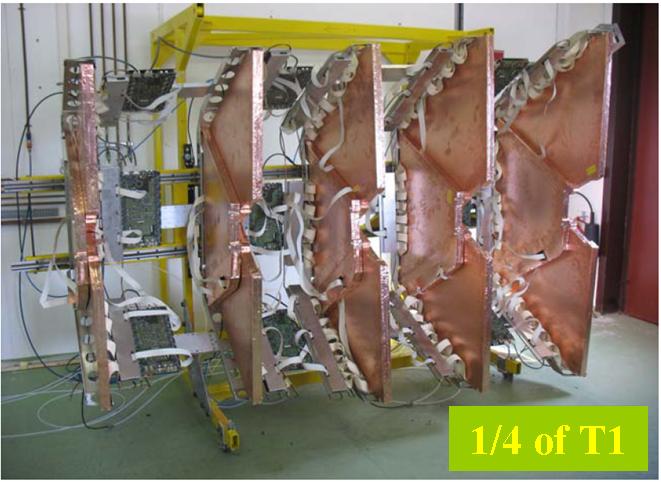}
\includegraphics[width=0.49\linewidth,height = 3.5 cm]{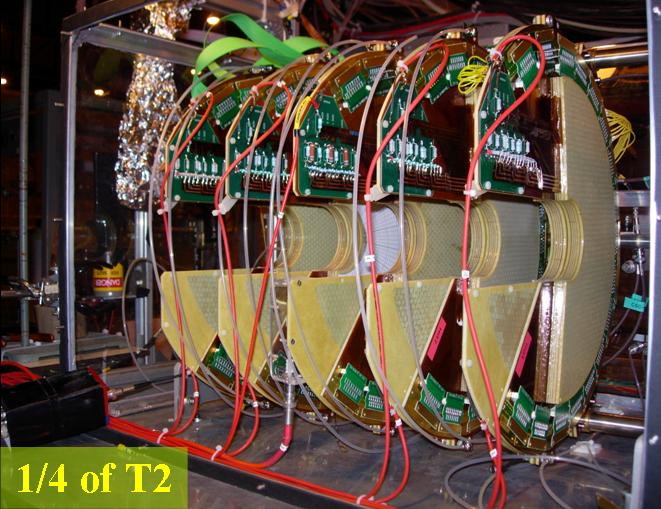}
\caption{Left (right): one half-arm of the T1 (T2) telescope.}
\label{fig:totem_T1T2}
\end{figure}

Charged track reconstruction in the 3.1 $<$ $|\eta|$ $<$ 6.5 range 
is provided by the T1 and T2 gas detectors, 
with a 2$\pi$ coverage and with a very high efficiency \cite{TOTEM_JINST}.
They allow the measurement of inelastic rates with small losses, given
their trigger capability with a geometric acceptance grater than 95$\%$ for all inelastic events.
At the same time they can provide the reconstruction of the interaction vertex, allowing 
to reject background events.
\newline
Each arm of the T1 telescope, located at $\sim$ 9 m from IP 5, covers the range 
3.1 $<$ $|\eta|$ $<$ 4.7 and is composed of five planes of ``Cathode Strip 
Chambers'' (CSC)~\cite{TOTEM_TDR} (figure~\ref{fig:totem_T1T2}, left).
Each plane consists of six trapezoidal CSCs, with 10 mm thick gas gap and a gas mixture of 
Ar/CO$_2$/CF$_4$ ($40\%/50\%/10\%$), providing the measurement of the charged particle 
hit position with a spatial resolution of $\sim$ 1 mm. The anode wires (pitch of 3 mm), also giving 
level-1 trigger information, are parallel to the trapezoid base, while the cathode strips (pitch of 
5 mm) are rotated by $\pm$ $60^o$ with respect to the wires.
\newline
The T2 telescope extends charged track reconstruction to the range 5.3 $<$ $|\eta|$ $<$
6.5 \cite{TOTEM_TDR}. It is  based on the ``Gas Electron Multiplier'' (GEM) technology 
\cite{GEM}, which has been chosen thanks to its good spatial resolution, excellent 
rate capability and good resistance to radiation.
Each T2 half-arm, located at $\sim$ 13.5 m from IP 5, is made by the combination of ten aligned 
detectors planes having an almost semicircular shape (figure~\ref{fig:totem_T1T2}, right). 
The T2 GEMs are characterized by a triple-GEM structure with a
gas mixture of Ar/CO$_2$ (70$\%$/30$\%$) \cite{TOTEM_GEM}.
The read-out board has two separate layers with different patterns: one with 256x2 concentric
circular strips (80\,$\mu$m wide, pitch of 400\,$\mu$m), allowing the track radial coordinate
reconstruction with a resolution of $\sim$ 100\,$\mu$m; the other with a
matrix of 24x65 pads (from 2x2\,mm$^2$ to 7x7\,mm$^2$ in size) providing the reconstruction of 
the track azimuthal coordinate with a resolution of $\sim$ 1$^o$, as well as level-1 trigger information.

\begin{figure}[htb]
\centering
\includegraphics[width=0.85\linewidth]{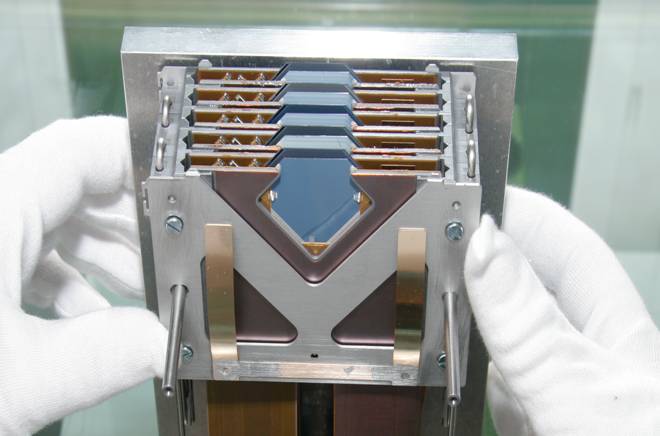}
\caption{Silicon detectors hosted in one pot.}
\label{fig:totem_RP}
\end{figure}

'`Leading'' protons with a scattering angle down to few $\mu$rad are detected by 
special movable beam-pipe insertions called Roman Pots (RPs).
They host silicon detectors which are moved very close to the beam, when it is in stable conditions.
As the detection of protons at such small scattering angles requires a
detector active area as close to the beam as $\sim$ 1 mm, a novel ``edgeless planar silicon''
detector technology has been developed for TOTEM RPs in order to have an edge dead zone minimized
to only $\sim$ 50 $\mu$m \cite{RP_Silicon}.
Each RP station is composed of two units, in order to have a lever harm for a better local track
reconstruction and a higher efficiency of trigger selection by track angle. Each unit
consists of three pots, two vertical and one horizontal completing the acceptance for diffractively 
scattered protons. Each pot contains a stack of 10 planes of silicon strip detectors 
(figure~\ref{fig:totem_RP}). Each plane has 512 strips (pitch of 66 $\mu$m), oriented at 
$+45^o$ (5 "$u$"-planes) or at $-45^o$ (5 "$v$"-planes) w.r.t. the detector edge facing 
the beam, allowing a single hit resolution of $\sim$ 10 $\mu$m.

The VFAT digital chip \cite{TOTEM_JINST}, specifically designed for TOTEM and characterized by 
trigger capabilities, provides the read-out of all TOTEM sub-detectors and the related trigger 
information. 
\section{Measurements at $\sqrt{s}$ = 7 TeV}
\label{7TeV_meas}
A precise measurement of small scattering angles for the protons (needed to access small 
|$t$| values) requires, besides a close approach of the detectors to the outgoing beams, 
the beam angular divergence (${\sigma}_{\Theta}^*$) to be as small as possible, 
at the level of few $\mu$rads. Since ${\sigma}_{\Theta}^*$ = $\sqrt{{\epsilon}/{\beta}^*}$, 
this can be obtained either by increasing the value of the betatron function $\beta$ at IP
(${\beta}^*$) or by reducing the beam emittance ($\epsilon$). 
Given the higher control on $\beta^*$, special LHC runs with a 
dedicated machine optics configuration at high $\beta^*$ are required. 
The consequent increase in beam size at the interaction point characterizes
these runs with a low luminosity ($\mathcal{L}$), which is anyway needed 
in TOTEM analyses in order to avoid extra interactions between the colliding bunches.
The data taken during 2010 and 2011 in special runs with various LHC configurations allowed the 
first TOTEM measurements, as reported in the following. 
\subsection{Elastic scattering}
\label{elas_scatt}
TOTEM has performed a first measurement of $d{\sigma}_{el} / dt$ in the
0.36 $<$ $|t|$ $<$ 2.5 GeV$^2$ range using data taken in 2010 with the standard LHC optics
($\beta^*$ = 3.5 m) during a dedicated run at low luminosity \cite{EPL95_SigEl}. A total
luminosity of 6.1 nb$^{-1}$ was integrated, with the RP detectors approaching the beams as 
close as 7 times the transverse beam size (${\sigma}_{b}$).
This measurement has been extended to |$t$| values as low as $2{\cdot}10^{-2}$ GeV$^2$ by 
analyzing the data recorded in a short run
taken in June 2011 with a dedicated beam optics configuration 
($\beta^*$ = 90 m) and low luminosity, during which a luminosity of 1.7 $\mu$b$^{-1}$ was 
integrated with the RPs placed at 10 ${\sigma}_{b}$ \cite{EPL96_SigEl_SigTOT}. 
This made for the first time the extrapolation to the optical point ($t$ = 0) possible, so 
allowing the determination of the elastic scattering cross-section (${\sigma}_{el}$) as well 
as of ${\sigma}_{tot}$ via the optical theorem. 
Even smaller |$t$| values, down to $5{\cdot}10^{-3}$ GeV$^2$, were then obtained with an updated 
analysis performed using different higher statistics data sets (for a total of 83 $\mu$b$^{-1}$) 
taken in October 2011 with special runs at $\beta^*$ = 90 m, where the RPs were put very close to 
the beam (4.8 to 6.5 ${\sigma}_{b}$) \cite{EPLXX_SigEl_SigTOT}. This enabled the 
observation of 91$\%$ of ${\sigma}_{el}$, to be compared to only 67$\%$ of the previous 
measurement, which gave an improved extrapolation to the optical point.
The data taking and analysis strategies were substantially the same 
for all these measurements. 

The background was significantly reduced a the trigger level by requiring trigger tracks 
on both sides of the IP.
A high trigger efficiency (over 99 $\%$ per proton) was achieved by using a loose trigger 
requiring a track segment in any of the vertical RPs of the 220 m stations. 
Elastic candidates were then selected offline by requiring a reconstructed track in 
both ($u$- and $v$-) projections of the vertical RP units on each side of the IP in a ``diagonal'' 
topology: top (bottom) left of IP - bottom (top) right of IP.

Using the ``optical functions'', which describe the explicit proton path from the IP to the RPs 
location through the LHC magnet elements as a function of the proton position ($x^*$, $y^*$, $z^*$ = 0) 
and scattering angle (${\Theta}_x^*$, ${\Theta}_y^*$) at the IP, the horizontal (${\Theta}_x^*$) and 
vertical (${\Theta}_y^*$) projections of the scattering angle are deduced from the measurement of the 
proton position at the RPs ($x$, $y$) according to the following equations:
\begin{equation}
 x = L_{x}{\Theta}_x^* + v_{x}x^*, \hspace{1.0cm} 
 y = L_{y}{\Theta}_y^* + v_{y}y^*, 
 \label{ProtTranEq}
\end{equation}
where $L_{x,y}$ and $v_{x,y}$ represent the optical functions at the RPs position. 
Being related to the betatron function, they depend on the LHC optics configuration and 
define the related $|t|$-range acceptance of the RPs. With the standard (${\beta}^*$ = 3.5 m) 
and 
intermediate (${\beta}^*$ = 90 m) optics the vertical scattering angle ${\Theta}_y^*$ can be directly 
reconstructed from the track position $y$, whereas the horizontal component ${\Theta}_x^*$ is 
optimally reconstructed from the local track angle ${\Theta}_x$ = $dx/ds$ at the RPs.

Dedicated procedures have been performed in order to ensure the precision and the reproducibility of
all RP detector planes alignment with respect to each other and to the position of the beam center,
one of the most delicate ad difficult tasks of the experiment. A precise relative alignment (at the level of few 
$\mu$m) of all three RPs in a unit has been obtained during the data analysis by correlating their position via 
common particle tracks reconstruction in the overlap zone of the horizontal RPs with the vertical ones. The global 
symmetrical alignment of all the RPs with respect to the beam center has been obtained (with a precision of $\sim$ 20 
$\mu$m) during dedicated beam fills by moving them towards the sharp beam edge cut by the beam collimators, until 
a beam losses spike is observed downstream of the RPs. A final horizontal and vertical alignment fine-tuning has 
then been achieved from studies on the reconstructed tracks by exploiting the azimuthal symmetry of elastic 
scattering. 

Given the collinearity of the elastically scattered protons the event selection has been performed by 
requiring a strict correlation (consistent with the beam divergence) between the two ${\Theta}_x^*$ (and 
${\Theta}_y^*$) projections, expected to be the same on both sides of IP.
Other cuts on the reconstructed kinematic variables have also 
been applied in order to remove diffractive (low $\xi$ cuts) and beam halo (vertex cuts) 
background. Acceptance limitations have been accounted for 
by assuming azimuthal symmetry of the elastic scattering and by correcting for smearing around the 
detector edges. Bin migration effects, due to angular resolution (related to detector resolution 
and beam divergence effects), have been corrected using a Monte Carlo (MC) based iterative procedure.
The RPs trigger efficiency ($>$ 99.8$\%$ @ 68$\%$ CL) was evaluated by using the ``zero-bias'' (trigger 
on bunch crossing) data stream. 
DAQ and reconstruction inefficiencies were calculated from dedicated studies on data.
The total luminosity associated to the collected data has been derived from the instantaneous
luminosity measured by CMS with an uncertainty of 4$\%$,
integrated over the data taking period. 
The systematic uncertainty was found to be dominated by the luminosity uncertainty 
for |$t$| < 0.2 GeV$^2$ and by t-dependent (optics-related) contributions elsewere.
\begin{figure}[!htb]
\centering
\includegraphics[width=1.\linewidth,clip]{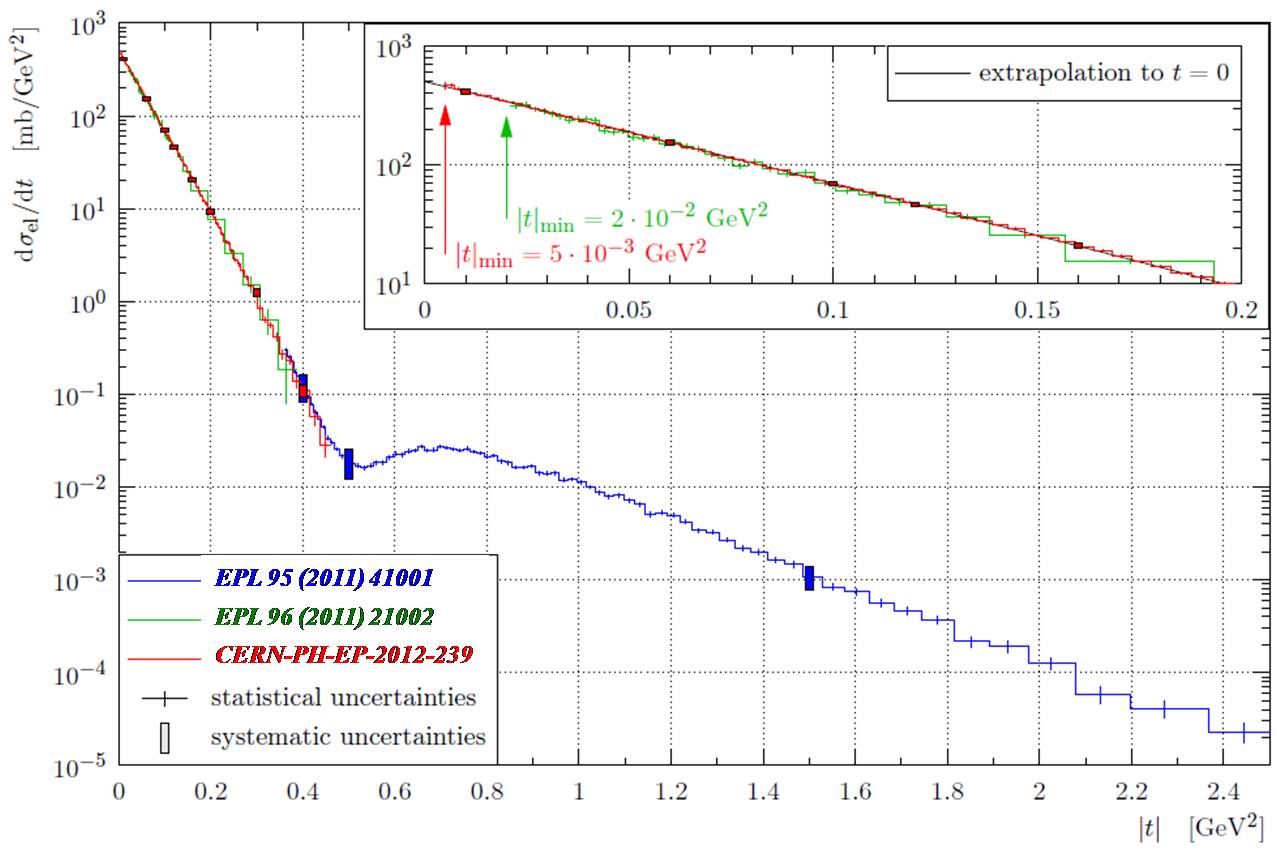}
\caption{$d{\sigma}_{el}/dt$ measurements performed by TOTEM in three 
different run conditions and RPs approaches to the beam. Each measurement is 
shown in a different color. On the right-top a zoom of the low $|t|$ region used for the 
extrapolation to $t$ = 0 is shown.
}
\label{fig:dSigmaEl_dt}
\end{figure}

All TOTEM differential cross-section measurements are given in figure~\ref{fig:dSigmaEl_dt}, 
where the right-top insert shows a zoom of the region used for the extrapolation to $t$ = 0 
in the two measurements at ${\beta}^*$ = 90 m.
For $|t|$ $<$ 0.2 GeV$^{2}$ the data can be described very well by an exponential function 
with slope $B = (19.9 \pm 0.3)$ GeV$^{-2}$ which, compared to previous collider experiments a lower energies, 
shows a steadily increasing value with the collision energy $\sqrt{s}$. 
The expected diffractive minimum, typically pronounced in pp scattering, is then observed at $|t| = (0.53 \pm 
0.01)$ GeV$^{2}$ and it is closer to $t$ = 0 respect to the 
measurements performed at lower energies. At higher $|t|$ the $d{\sigma}_{el} / dt$ can be described by a 
power law $|t|^{-n}$ with $n = 7.8 \pm 0.3$, in the $|t|$-range 1.5 - 2.0 GeV$^2$. 
The comparison with the predictions of different available theoretical models shows a partial consistency 
with the data (slope $B$, dip position and exponent $n$ at large $|t|$) only for some of them \cite{EPL95_SigEl}.

The low $|t|$ values reached with the ${\beta}^*$ = 90 m optics made the exponential extrapolation to $t$ = 0 
($\left.\frac{d{\sigma}_{el}}{dt} \right|_{t=0}$) possible, allowing to derive (by integration of the elastic 
distribution) the elastic cross section. In particular, the measurement performed at lower $|t|$ (down to 
$5{\cdot}10^{-3}$ GeV$^2$) gave the most precise extrapolation to the optical point, 
so the best ${\sigma}_{el}$ measurement. The ${\sigma}_{el}$ results obtained with the two different 
data sets are reported in table~\ref{tab-1}.

The optical theorem, relating the total to the elastic pp cross-section,  
was then used to obtain ${\sigma}_{tot}$ and the inelastic cross-section (${\sigma}_{inel}$) according to:
\begin{equation}
 \sigma_{tot}^{2} = \frac{16{\pi}{({\hbar}c)}^2}{1 + \rho^{2}} \cdot
 \left.\frac{d{\sigma}_{el}}{dt} \right|_{t=0}, \hspace{0.5cm}  
 {\sigma}_{inel} = {\sigma}_{tot} - {\sigma}_{el},
 \label{eqn_sigma_t_fromEl}
\end{equation}
where for the $\rho$ parameter\footnote{
 $\rho = {\Re}|f_{el}(0)|/{\Im}|f_{el}(0)|$, where $f_{el}(0)$ is the forward 
 nuclear elastic amplitude. 
} 
was chosen the COMPETE preferred-model extrapolation of 0.141 $\pm$ 0.007 \cite{COMPETE}.  
The errors on this measurement are dominated by the uncertainties related to the luminosity
and to the extrapolation to $t$ = 0.
The results obtained with the two different data sets are summarized in table~\ref{tab-1}. 
\subsection{Inelastic cross section}
\label{inelas_xsec}
Based on the same data taking as in reference \cite{EPLXX_SigEl_SigTOT} ${\sigma}_{inel}$ 
was also directly measured using inelastic events triggered by the T2 
telescope and complemented by T1 telescope data, with the luminosity 
determination being provided by CMS \cite{EPLXX_Sig_Inel}. 

In order to perform an optimized study of the trigger efficiency and of the 
beam-gas background corrections, the inelastic events were separated into 
three subsamples: with tracks in both T2 arms (mainly due to non diffractive and to double 
diffractive processes) and with tracks in only one (``+'' or ``-'') of the two T2 arms 
(mainly due to single diffractive processes). 
Special emphasis was given to properly tune the simulation of the forward region, 
in particular in front and around T2, since the secondary particles, produced in this zone by 
the interaction of primary particles with the material, enhance the observed inelastic 
rate. To obtain the true inelastic rate the observed one had to be properly corrected. 
The corrections, derived with studies on proper MC samples, ``zero bias'' and 
``non colliding bunch'' trigger data, 
were done in three steps allowing to get: the ``T2 visible'' cross-section 
($\sim$ 95$\%$ of ${\sigma}_{inel}$, related to events with at least one particle 
in the $|\eta|$ acceptance of T2, i.e. 
with 5.3 < $|\eta|$ < 6.5); the ``global visible'' cross-section (related to events with 
at least one particle up to the $|\eta|$ acceptance of T2, i.e. with $|\eta|$ < 6.5); 
the ``missing'' cross-section (related to events with only particles beyond the T2 acceptance, 
i.e. with $|\eta|$ > 6.5, mainly due to single diffractive processes with masses below 
$\sim$ 3.4 GeV/c$^2$). 
These studies showed in particular a very good performance of the T1 and T2 detectors, 
characterized by an event reconstruction efficiency of about 98$\%$ (99$\%$) for an individual 
T1 (T2) arm. 
After comparing several MC models, the correction for the missing low mass single diffractive 
cross-section has been obtained based on the QGSJET-II-03 generator, which can  well 
describe the measurements of low mass diffraction at lower energies. This correction 
(4.2 $\pm$ 2.1 $\%$) is the largest one to the inelastic rate and, given the large uncertainty, 
represents the second largest systematic uncertainty of this measurement after the 
one related to the luminosity (4$\%$). 
The result, reported in table~\ref{tab-1}, is well in agreement with the previous TOTEM 
measurements deduced from the elastic scattering via the optical theorem \cite{EPL96_SigEl_SigTOT,  
EPLXX_SigEl_SigTOT} as well as compatible with the measurements performed by ALICE \cite{SigInel_ALICE}, 
ATLAS \cite{SigInel_ATLAS} and CMS \cite{SigInel_CMS}.   
However, given the unique $\eta$ coverage of the T1 and T2 detectors allowing to reach the lowest 
diffractive masses (M$_X$ > 3.4 GeV/c$^2$), this measurement is characterized by a minimal model 
dependence for the required low mass diffraction corrections respect to the measurements 
performed by the other LHC experiments. 

Furthermore, an estimate of low mass diffraction contribution was obtained by comparing the full 
inelastic cross-section measurement of reference \cite{EPLXX_SigEl_SigTOT} 
(by construction derived without any assumption about low mass diffraction), 
with the direct ``global visible'' ($|\eta|$ < 6.5, or M$_X$ > 3.4 GeV/c$^2$) inelastic cross-section measurement. 
From their difference a ``missing'' inelastic cross-section of 2.6 $\pm$ 2.2 mb was obtained, 
corresponding to an upper limit of 6.3 mb at 95$\%$ CL for diffractive events 
with M$_X$ < 3.4 GeV/c$^2$.     
\subsection{Luminosity independent total cross-section}
\label{total_xsec}
A ``$\rho$-independent'' determination of ${\sigma}_{tot}$ has been obtained by summing
directly the elastic \cite{EPLXX_SigEl_SigTOT} and the inelastic \cite{EPLXX_Sig_Inel} cross sections: 
\begin{equation}
 {\sigma}_{tot} = {\sigma}_{el} + {\sigma}_{inel}
 \label{eqn_sigmatot_sum}
\end{equation}
This result, reported in table~\ref{tab-1}, is characterized by a larger uncertainty 
due to the direct propagation of the luminosity uncertainty.

\begin{figure}[!htb]
\centering
\vspace{-0.55cm}
\includegraphics[width=1.05\linewidth,clip]{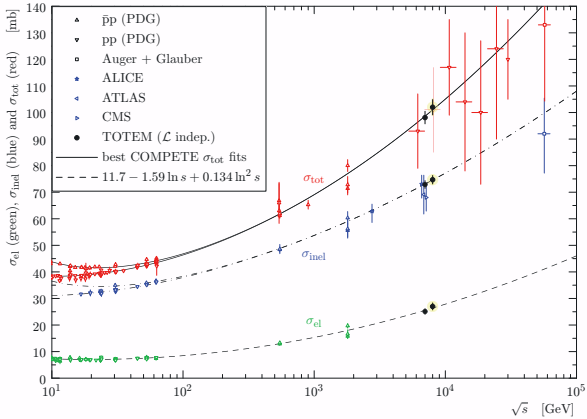}
\vspace{-1.3cm}
\caption{Compilation of the ${\sigma}_{tot}$, ${\sigma}_{inel}$ and ${\sigma}_{el}$ measurements.
The black points show the $\mathcal{L}$-independent measurements performed by TOTEM 
at $\sqrt{s}$ = 7 and $\sqrt{s}$ = 8 TeV. The continuous black lines (lower for $pp$, upper for $p\bar p$) 
represent the best fit to ${\sigma}_{tot}$ data by the COMPETE Collaboration \cite{COMPETE}, while the 
dashed line is related to a fit to the ${\sigma}_{el}$ data. The dash-dotted lines, corresponding to 
${\sigma}_{inel}$, are derived as the difference between the two previous fits. 
}
\label{fig:sigma_tot}
\end{figure}

However, the dependence of the ${\sigma}_{tot}$ measurement on the luminosity can be eliminated 
using the optical theorem:
\begin{equation}
 \sigma_{tot} = \frac{16{\pi}({\hbar}c)^2}{1 + \rho^{2}} \cdot
 \frac{dN_{el}/dt |_{t=0}}{N_{el} + N_{inel}},
  \label{eqn_sigmatot}\\
\end{equation}
where $N_{el}$ and $N_{inel}$ represent respectively the elastic and inelastic rates integrated over a given 
data taking period. 
Using the above equation, the elastic and inelastic measurements performed (independently but simultaneously) 
in the same analyses of reference \cite{EPLXX_SigEl_SigTOT} and \cite{EPLXX_Sig_Inel} respectively, 
were combined to obtain ${\sigma}_{tot}$ without taking into account the 
luminosity \cite{EPLXX_SigEl_TOT_Inel_Lindep_7TeV}.  
Furthermore, also ${\sigma}_{el}$ and ${\sigma}_{inel}$ were derived independently from 
the luminosity using the measured ratio $N_{el}$/$N_{inel}$. 
The errors on this measurement are dominated by the uncertainties on the extrapolation to $t$ = 0 and on the 
correction for the contribution from low mass diffraction.
The cross-sections measured with this ``$\mathcal{L}$-independent'' method, reported in table~\ref{tab-1}, 
are well in agreement with the previous TOTEM measurements, which confirms the understanding of the systematic 
uncertainties and of the corrections applied in the different analysis strategies. 
\begin{table}
\centering
\caption{Summary of TOTEM elastic, inelastic and total cross-section 
measurements at $\sqrt{s}$ = 7 TeV. The first two lines refer to the ``elastic only'' 
measurements (eq.~\ref{eqn_sigma_t_fromEl}) related to the June 
2011 \cite{EPL96_SigEl_SigTOT} 
and October 2011 \cite{EPLXX_SigEl_SigTOT} data taking periods. The third and fourth lines report 
respectively the $\rho$-independent (eq.~\ref{eqn_sigmatot_sum}) \cite{EPLXX_SigEl_SigTOT,EPLXX_Sig_Inel} and 
the $\mathcal{L}$-independent (eq.~\ref{eqn_sigmatot}) \cite{EPLXX_SigEl_TOT_Inel_Lindep_7TeV} 
measurements for the October 2011 data.
}
\label{tab-1}       
\begin{tabular}{llll}
\hline
Data analysis                                   & $\sigma_{el}$ (mb) & $\sigma_{inel}$ (mb) & $\sigma_{tot}$ (mb) \\
\hline
\cite{EPL96_SigEl_SigTOT}                       & 24.8 $\pm$ 1.2     & 73.5 $\pm$ 1.6       & 98.3 $\pm$ 2.8 \\
\cite{EPLXX_SigEl_SigTOT}                       & 25.4 $\pm$ 1.1     & 73.2 $\pm$ 1.3       & 98.6 $\pm$ 2.2 \\
\cite{EPLXX_SigEl_SigTOT} \cite{EPLXX_Sig_Inel} & 25.4 $\pm$ 1.1     & 73.7 $\pm$ 3.4       & 99.1 $\pm$ 4.3 \\
\cite{EPLXX_SigEl_TOT_Inel_Lindep_7TeV}         & 25.1 $\pm$ 1.1     & 72.9 $\pm$ 1.5       & 98.0 $\pm$ 2.5 \\
\hline
\end{tabular}
\end{table}
These measurements are also reported in figure~\ref{fig:sigma_tot}, showing a very good agreement with the
expectations from the overall fit of previously measured data over a large range of energies \cite{COMPETE}.

The optical theorem can also be applied in a complementary way in order to obtain the 
integrated luminosity (${\mathcal{L}}_{int}$) for a given data taking period:
\begin{equation}
 \mathcal{L}_{int} = \frac{1 + \rho^{2}}{16{\pi}({\hbar}c)^2} \cdot
 \frac{(N_{el} + N_{inel})^{2}}{dN_{el}/dt |_{t=0}}
  \label{eqn_lumi}
\end{equation}
Thus, the elastic and inelastic rates were also combined to obtain an absolute 
(and independent) ${\mathcal{L}}_{int}$ measurement. The TOTEM results were found to be well in agreement 
with the CMS ones within their uncertainties ($\sim$ 4$\%$ for both) \cite{EPLXX_SigEl_TOT_Inel_Lindep_7TeV}.
\subsection{Forward charged particle pseudorapidity density}
\label{dN_dEta}
A measurement of the pseudorapidity density (dN$_{ch}$/d$\eta$) of forward charged particles
was also performed by TOTEM using the data taken in May 2011 with an inclusive
T2 trigger during special LHC fills at low luminosity \cite{EPL96_dN_dEta}.

As a key aspect of this study is an optimal particle track reconstruction, 
a particular effort has been devoted in the analysis in order to correct for misalignment biases, 
found to be dominated by global displacements of the T2 quarters.
The relative positions of the detector planes within a T2 quarter in terms of $x$- and $y$-shifts (internal 
alignment) have been obtained using two different methods (iterative and MILLEPEDE), giving consistent 
results in the related corrections with an uncertainty of about 30 $\mu$m. 
The relative alignment between the two neighbouring quarters of an arm has been derived using tracks
reconstructed in the overlap regions.
The global detector alignment respect to its nominal position (corrections for tilts and shifts), 
of main importance for the present analysis, has been achieved
by exploiting the expected symmetry in the track parameters distributions,
as well as the position on each T2 plane of the ``beam pipe shadow'' (a circular-shaped zone 
characterized by a very low track efficiency due to primary particles interaction in the 
$\eta \sim $ 5.54 beam pipe cone in front of T2). The combination of all these methods 
resulted in corrections for $x$- and $y$-shifts with a precision of $\sim$ 1 mm and for 
$xz$- and $yz$-plane tilts with a precision of $\sim$ 0.4 mrad.  

Secondary track rejection is another important step of this analysis as about 80$\%$ 
of the T2 reconstructed tracks is due to secondaries. It has been performed with a procedure for the 
primary/secondary particle discrimination based on a proper track parameter, which has been found 
in detailed MC studies to be the most effective for this task and the most stable against misalignment errors.
Primary track efficiency (found to be $\sim$ 80$\%$) and smearing effects corrections have been 
obtained from MC studies. 

\begin{figure}[!htb]
\centering
\vspace{-0.3cm}
\includegraphics[width=1.\linewidth,clip]{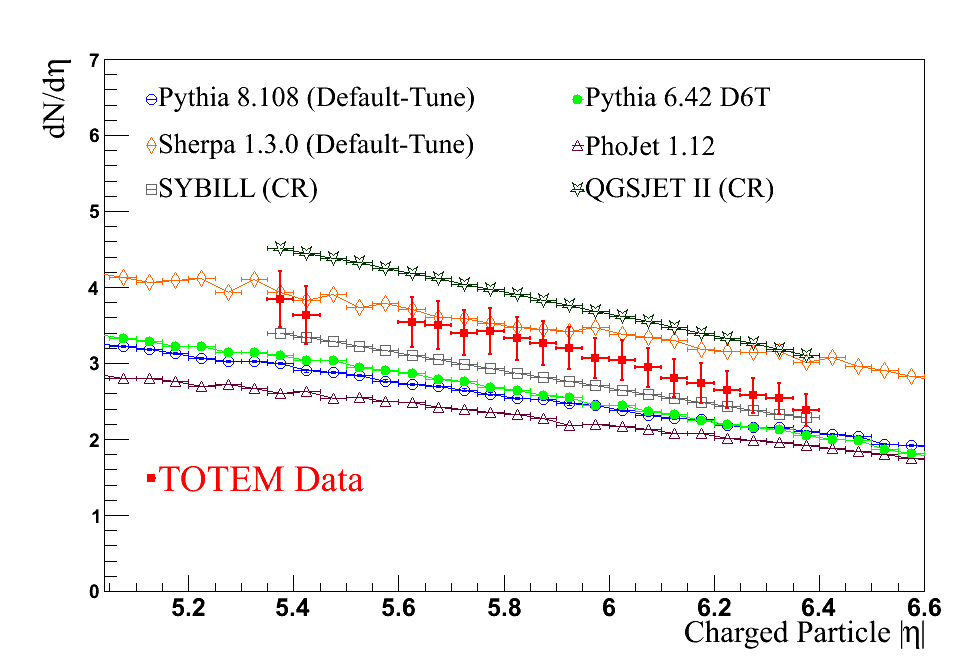}
\vspace{-0.7cm}
\caption{dN$_{ch}$/d$\eta$ distribution measured by Totem. The experimental results (red 
squares), obtained as the average of the four T2 quarters, are reported with the 
error bars including both statistical and systematic errors. The predictions from some 
MC generators are reported for comparison.  
}
\label{fig:dN_dEta}
\end{figure}

The results of the measurement, which refers to events with at least one 
charged particle with $P_T$ $\ge$ 40 MeV/c and $\tau$ $>$ 
$0.3{\cdot}10^{-10}$ s going in the T2 acceptance region, 
are reported in figure~\ref{fig:dN_dEta}. Here the red squares show the experimental 
values in terms of the average of the results found with the four T2 quarters, with the error 
bars representing the combination of the statistical and systematic errors (up to $\sim$ 10$\%$), 
dominated by primary track efficiency and global alignment corrections uncertainties.
Figure~\ref{fig:dN_dEta} also shows the comparison of some MC expectations with the TOTEM results.
None of the theoretical models has been found to fully describe the data, the cosmic ray MC generators
(SYBILL and QGSJET-II) showing a better agreement for the slope.
\section{Measurements at $\sqrt{s}$ = 8 TeV}
\label{8TeV_meas}
TOTEM also took data at $\sqrt{s}$ = 8 TeV during 2012 in special LHC runs. In some of them the 
first common data taking with CMS was also performed, with a TOTEM $\Leftrightarrow$ CMS trigger 
exchange. The October 2012 run was in particular characterized by ${\beta}^*$ = 1000 m and a 
RPs approach of 3 ${\sigma}_{b}$ to the beam center, allowing to measure ${\sigma}_{el}/{dt}$ down 
to $6{\cdot}10^{-4}$ GeV$^2$ so in principle opening the possibility of a determination of the 
$\rho$ parameter by studying the Coulomb-Nuclear interference region. 
Several analyses on these data are at the moment ongoing.

The data samples recorded on July 2012 during special fills with ${\beta}^*$ = 90 m have been used to 
perform the $\mathcal{L}$-independent measurement of ${\sigma}_{tot}$ (101.7 $\pm$ 2.9 mb), 
${\sigma}_{el}$ (27.1 $\pm$ 1.4 mb) and ${\sigma}_{inel}$ (74.7 $\pm$ 1.7 mb) at $\sqrt{s}$ = 8 TeV 
\cite{PRL_SigEl_TOT_Inel_Lindep_8TeV}. These results, also reported in figure~\ref{fig:sigma_tot}, 
are again in very good agreement with the expectations from the overall fit of previously 
measured data. 
\section{Summary and Conclusions}

Using the data taken during 2010 and 2011 in special LHC runs with different beam 
conditions ($\beta^* =$ 3.5 and 90 m, low $\mathcal{L}$) the TOTEM experiment has performed its first 
measurements at $\sqrt{s}$ = 7 TeV. These comprise the $d{\sigma}_{el}/dt$ 
measurement in a wide |$t$| range ($5{\cdot}10^{-3}$ $<$ $|t|$ $<$ 2.5 
GeV$^2$), as well as the measurement of ${\sigma}_{el}$, ${\sigma}_{inel}$ and 
${\sigma}_{tot}$ 
using different methods. These last results are in good agreement within the errors with the theoretical 
expectations and, for what concerns ${\sigma}_{inel}$, with the measurements of other LHC experiments. 
The dN$_{ch}$/d$\eta$ distribution was also measured in the 5.3 $<$ |$\eta$| $<$ 6.4 range.
\newline
The analysis of the data taken in 2012 at $\sqrt{s}$ = 8 TeV with ${\beta}^*$ = 90 m gave the 
$\mathcal{L}$-independent measurement of ${\sigma}_{tot}$, ${\sigma}_{el}$ and ${\sigma}_{inel}$, 
again found in good agreement with the theoretical predictions.
Work is in progress for the analysis of other data samples taken at $\sqrt{s}$ = 8 TeV. In particular,   
the ones with $\beta^* =$ 1000 m are expected to open the possibility of the $\rho$ parameter measurement,  
while the ones taken jointly with CMS will provide the first common physics results on hard and soft diffraction as 
well as the dN$_{ch}$/d$\eta$ measurement over the whole $\eta$ range up to $|\eta|$ $\sim$ 6.5.
\section{Acknowledgements}
We gratefully thanks the Conference Organizers for their kind 
invitation to this very nice event in the wonderful town of Kyoto.
\vspace{-7.0cm}
\end{document}